\documentclass[sigconf, nonacm]{acmart}
\AtBeginDocument{%
  }

\usepackage{booktabs}
\usepackage{array}
\usepackage{multirow}
\usepackage{pifont}
\usepackage{pgfplots}
\pgfplotsset{compat=1.17}
\usetikzlibrary{positioning,arrows.meta}

\newcommand{\up}[1]{\textbf{+#1}}
\newcommand{\dn}[1]{$-$#1}

\usepackage{xurl}  
\begin{document}

\title{SkillRep: Characterizing and Understanding the Representation Design Space of Agent Skill Retrieval}

\author{Tianyi Sun}
\authornote{Equal contribution.}
\affiliation{\institution{EverMind, Shanda Group}\institution{Peking University}\country{China}}
\email{tianyi.sun@shanda.com}

\author{Pengfei Yao}
\authornotemark[1]
\authornotemark[3]
\affiliation{\institution{EverMind, Shanda Group}\country{China}}
\email{yaopengfei@shanda.com}

\author{Yanze Wang}
\affiliation{\institution{EverMind, Shanda Group}\institution{Peking University}\country{China}}
\email{yanze.wang@shanda.com}

\author{Hongda Chen}
\affiliation{\institution{EverMind, Shanda Group}\institution{Southeast University}\country{China}}
\email{hongda.chen@evermind.ai}

\author{Yunyun Han}
\affiliation{\institution{EverMind, Shanda Group}\country{China}}
\email{hanyunyun@shanda.com}

\author{Jun Sun}
\authornote{Corresponding authors.}
\affiliation{\institution{Peking University}\country{China}}
\email{jsun@pku.edu.cn}

\author{Chuanrui Hu}
\authornotemark[2]
\authornote{Project leaders.}
\affiliation{\institution{EverMind, Shanda Group}\country{China}}
\email{chuanrui.hu@shanda.com}

\author{Yafeng Deng}
\authornotemark[2]
\affiliation{\institution{EverMind, Shanda Group}\country{China}}
\email{dengyafeng@shanda.com}

\renewcommand{\shortauthors}{Sun et al.}

\begin{abstract}
Retrieving the right skill from a large library is essential for LLM agents, yet requests and skills are written for different purposes and in different words. Existing retrievers narrow this request--skill gap with one or two representation choices each, so it is unknown whether their gains hold in combination. No prior work measures the gap itself, so it is also unclear which part of it each choice narrows. We define a representation design space by five choices---encoder fine-tuning, query-side and document-side expansion, background projection and a lexical channel---and evaluate them together for the first time, as modules of one retriever, SkillRep. With a 0.6B encoder and LLM expansion, SkillRep improves MRR@10 on SRA-Bench over the strongest same-size retriever by 7.02 points with the official skill pool and 8.79 with UniPool, a 111{,}247-skill pool we build from public registries, and is best on SkillsBench. The value of each choice depends on the others. Fine-tuning, for instance, reverses the ranking of the four training-free modules and turns them from complementary into overlapping. To understand why, we measure the gap with three quantities: \emph{background}, the embedding component that does not separate skills; \emph{unsaid vocabulary}, the gold skill's content a request leaves unsaid; and \emph{dilution}, the information lost when a skill is encoded into a single vector. They show which part each module narrows and by what mechanism. This reveals that modules narrowing the same part compete for it, while those narrowing different parts complement each other.
\end{abstract}



\maketitle

\section{Introduction}
\label{sec:intro}

\begin{figure*}[t]
  \centering
  \includegraphics[width=0.88\textwidth]{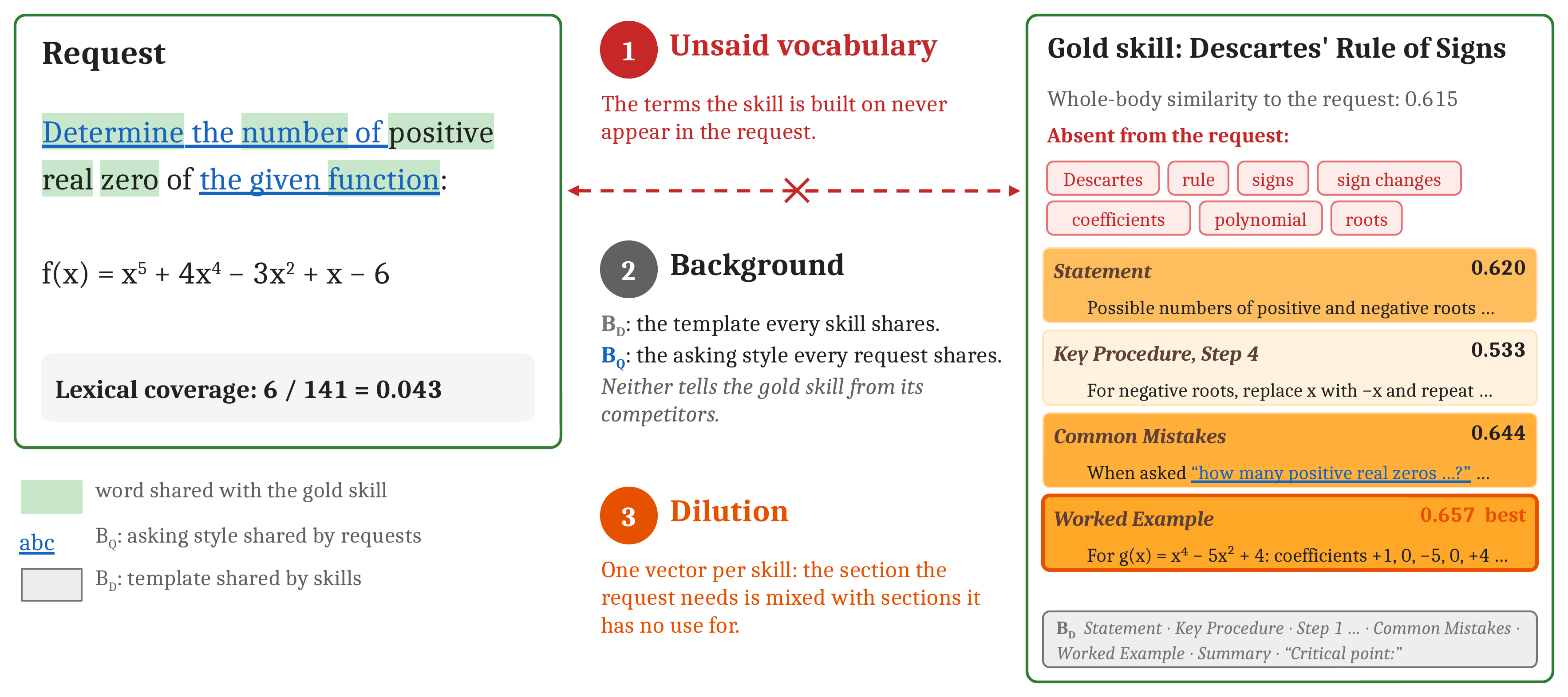}
  \caption{The three quantities on one SRA-Bench~\cite{su2026sra} request and its gold skill. The background is drawn schematically.}
  \label{fig:main}
\end{figure*}

LLM agents increasingly extend their capabilities with \emph{skills}~\cite{anthropic2025agentskills},
documents that describe how to carry out a type of task through procedures, tool usage and
examples. Skills are shared through public Web registries, which together already hold over a
hundred thousand of them~\cite{wang2026skillcorpus}, and retrieving a skill from these registries is a
Web search task performed by an agent rather than by a person. Since an agent cannot read every
skill for every task, it first retrieves a few candidates for the request at hand. If the right
skill is not among them, the agent cannot use
it~\cite{zheng2026skillrouter,sorathiya2026enrich}.

A request in SRA-Bench~\cite{su2026sra} reads \emph{``Determine the number of positive real zero of
the given function: $f(x)=x^5+4{\ast}x^4-3x^2+x-6$.''} The skill that answers it is
\texttt{Descartes' Rule of Signs}, described as \emph{``Counting possible positive
and negative real roots of a polynomial using sign changes in coefficients.''} The
request never names the rule, and the document never mentions this polynomial
(Figure~\ref{fig:main}). Existing methods narrow this distance by changing how the two
kinds of text are represented. Requests are rewritten toward skill
documents~\cite{liu2026skilldreamer,jiang2026skillreason}, documents toward
requests~\cite{ding2026skill2query,wang2026capability}, generic directions are projected
out~\cite{xiao2026skillsight}, and encoders are trained to cross the distance on their
own~\cite{jiang2026skillreason,chen2026adapting}. Each of these is a representation choice.
Prior work adopts one or two at a time, each evaluated on its own encoder and pool, so it is
unclear how much of a choice's gain survives a fine-tuned encoder or the other choices. Nor
does it measure the gap, so it cannot say which part a choice narrows.

Together these choices form a representation design space for skill retrieval. We combine
five of them in one retriever, SkillRep, which pairs a contrastively fine-tuned 0.6B encoder with four
training-free modules, namely query-side and document-side LLM expansion, background projection
and a BM25~\cite{robertson2009bm25} channel. On SRA-Bench, SkillRep improves MRR@10
over the strongest retriever of its size by $7.02$ points on the official pool and by $8.79$
on UniPool, a $111{,}247$-skill pool that we build from public registries, and it is best on
both SkillsBench~\cite{li2026skillsbench} tiers. The value of
each choice depends on the others. Fine-tuning, for instance, reverses the ranking of the four
training-free modules, so that the most valuable changes from background projection to query-side expansion, and
turns them from complementary into overlapping.

To understand these phenomena, we measure the gap with three per-request quantities. The
\emph{background} is a component shared by the embeddings of all skill documents, or of all
requests, so it does not separate the needed skill from its competitors. The \emph{unsaid
vocabulary} is the gold skill's content that a request leaves unsaid, which we measure by
\emph{coverage}, the share of the gold document's content words that the request expresses.
\emph{Dilution} is the information lost when a skill is encoded into a single vector, which
mixes the part that a request needs with all the other parts.

Tracing each module through the three quantities, we first identify which part of the gap
each one narrows (Table~\ref{tab:summary}). This explains how the modules
interact. Those that narrow the same part of the gap overlap, and those that narrow different
parts complement each other. Fine-tuning shrinks the documents'
background, so projecting it out adds much less, and it makes the
requests' background specific to requests, so that query-side expansion now narrows it as well
and overlaps with the other modules that do. We also uncover
the mechanism by which each module narrows its part of the gap. Query-side expansion helps through the gold document's words that
the LLM supplies. Document-side
expansion narrows the requests' background, and about half of its gain needs nothing
specific to a skill. Dilution hurts only the one-vector
representation, and the lexical channel recovers part of its cost by counting the words a
request shares with the gold document, which the document's single vector dilutes.

\paragraph{Contributions.}
\begin{itemize}
\item \textbf{Characterizing the representation design space}: encoder fine-tuning,
  query-side and document-side expansion, background projection and a lexical channel, five
  representation choices of which prior retrievers use one or two (Table~\ref{tab:related}). We are the first to combine all five in one retriever.
\item \textbf{Results and data}: SkillRep sets a new state of the art on SRA-Bench and
  SkillsBench with a 0.6B encoder, and we construct UniPool, a $111{,}247$-skill library from
  public registries with fewer near-duplicates than existing benchmark pools.
\item \textbf{Explaining these choices}: three per-request measures of the request--skill
  gap that show which part of the gap each module addresses and why modules that narrow the
  same part of the gap overlap.
\end{itemize}
Our code is available at \url{https://github.com/EverMind-AI/SkillCorpus/tree/main/skillrep}.

\section{Related Work}
\label{sec:related}

\paragraph{Skill retrieval.} As skill libraries outgrow what an agent can read in context, skill
retrieval has emerged as a research problem in its own right. Several benchmarks evaluate the task. SRA-Bench~\cite{su2026sra}
pairs task requests from six source datasets with a skill library,
SkillsBench~\cite{li2026skillsbench} measures how much skills help agents complete tasks, and
SkillRet~\cite{cho2026skillret} provides a large library with training and test splits. A survey
catalogs these and other benchmarks~\cite{ding2026skillevalevolution}. Other studies analyze the
setting itself, examining retrieval at registry scale~\cite{yang2026skillseek}, the quality of
published skills~\cite{zhang2026skillreusability}, and whether retrieved skills are compatible
with the request~\cite{wang2026skillisnotdocument}. On the method side, existing retrievers each
adopt one or two of the five representation choices (Table~\ref{tab:related}). One group
fine-tunes the encoder on request--skill pairs~\cite{zheng2026skillrouter,chen2026adapting,cho2026skillret}. SkillReason~\cite{jiang2026skillreason} additionally distills LLM reasoning about the request
into it. A second group narrows the gap with LLM-written text. SkillDreamer~\cite{liu2026skilldreamer}
expands the request by imagining how a capability would be realized, while other methods expand
the documents offline into pseudo-queries~\cite{ding2026skill2query}, capability
pages~\cite{wang2026capability}, enriched profiles~\cite{sorathiya2026enrich}, typed
pseudocode~\cite{li2026sap} or evidence layers~\cite{liang2026ssl}. A third group changes how
the existing documents are represented. SkillSight~\cite{xiao2026skillsight} projects out a
generic background subspace, field-aware retrieval~\cite{goulart2026fieldaware},
SkillFlow~\cite{li2025skillflow} and Skills in the Wild~\cite{liu2026wild} index different
fields and add a lexical channel, and SkillFeed segments skills by
section~\cite{wang2026skillfeed}. Beyond first-stage retrieval, further work reranks the
retrieved skills~\cite{chen2026skillreranker}, routes from a frozen LLM's hidden
states~\cite{chen2026routerwithin}, or compiles retrieved skills into compact
context~\cite{meng2026skillrae}. However, these choices have so far been evaluated only in isolation, each on its own encoder
and library, and never together.

\begin{table}[t]
    \caption{Representation choices each skill retriever uses: encoder fine-tuning (FT), query-side and document-side expansion (QE, DE), background projection (BP) and a lexical channel (Lex).}
  \label{tab:related}
  \small
    \setlength{\tabcolsep}{4pt}
  \newcommand{\y}{\ding{51}}
  \begin{tabular}{@{}lccccc@{}}
    \toprule
    Method & FT & QE & DE & BP & Lex\\
    \midrule
    SkillRouter~\cite{zheng2026skillrouter} & \y & & & & \\
    SkillReason~\cite{jiang2026skillreason} & \y & & & & \\
    Adapting Emb.~\cite{chen2026adapting} & \y & & & & \\
    SkillRet-Emb.~\cite{cho2026skillret} & \y & & & & \\
    SkillDreamer~\cite{liu2026skilldreamer} & & \y & & & \\
    Skill2Query~\cite{ding2026skill2query} & & & \y & & \\
    Capability Pages~\cite{wang2026capability} & & & \y & & \\
    Enrich-Retrieve-Rank~\cite{sorathiya2026enrich} & & & \y & & \y\\
    SkillSight~\cite{xiao2026skillsight} & & & & \y & \y\\
    Field-aware~\cite{goulart2026fieldaware} & & & & & \y\\
        SkillFlow~\cite{li2025skillflow} & & & & & \y\\
    Skills in the Wild~\cite{liu2026wild} & & & & & \y\\
    \midrule
    SkillRep & \y & \y & \y & \y & \y\\
    \bottomrule
  \end{tabular}
\end{table}

\paragraph{Anisotropy of embeddings.} Studies of text embeddings find that they are
an\-iso\-tropic, occupying a narrow cone of the space~\cite{ethayarajh2019contextual}, so a few
directions shared by all texts dominate their similarities and make them less informative.
Post-processing corrects this after training, by removing the mean and a few dominant
directions~\cite{mu2018allbutthetop}, by whitening the embeddings~\cite{su2021whitening}, or by
mapping them to an isotropic distribution with a normalizing flow~\cite{li2020bertflow}.
Contrastive learning instead reduces anisotropy during training, by pushing apart the
embeddings of unrelated texts~\cite{gao2021simcse}. A related effect appears in contrastive
multi-modal models, whose two modalities stay in separated regions of the space after training,
a phenomenon known as the modality gap~\cite{liang2022modalitygap}.

\paragraph{Techniques from information retrieval.} Several techniques from general information
retrieval address the mismatch between how queries and documents are worded. Query expansion
adds to a query the terms it lacks. With LLMs, HyDE~\cite{gao2023hyde} retrieves with the
embedding of a hypothetical answer document written for the query,
query2doc~\cite{wang2023query2doc} appends a generated pseudo-document to the query, and query
rewriting reformulates it~\cite{ma2023queryrewriting}. Document expansion works from the other
side, appending to each document the queries it could
answer~\cite{nogueira2019doc2query,nogueira2019doctttttquery}, optionally filtered for
quality~\cite{gospodinov2023doc2queryminus}. Such generated queries also serve to train
retrievers~\cite{bonifacio2022inpars,dai2023promptagator}. Hybrid retrieval combines dense
retrievers~\cite{karpukhin2020dpr,izacard2022contriever,ni2022gtr,wang2022e5}, which match by
meaning, with sparse ones such as BM25~\cite{robertson2009bm25}, which match exact
terms~\cite{ma2022anotherlookdpr}, and fuses their results by rank~\cite{cormack2009rrf} or by
score~\cite{fox1994combination,bruch2023fusion}. When such techniques are combined, however,
their improvements often fail to add
up~\cite{armstrong2009additivity,kharazmi2016additivity,akcay2017additivity}, and gains reported
over weak baselines often shrink over strong ones, for neural models as
well~\cite{lin2018hype,yang2019neuralhype}.

\section{SkillRep}
\label{sec:method}

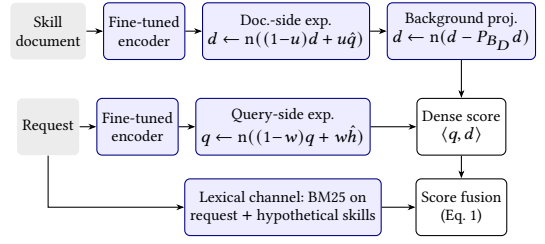
\begin{figure}[t]
  \centering
  \begin{tikzpicture}[font=\scriptsize, node distance=2.2mm and 2.2mm, >={Stealth[length=3pt]},
      box/.style={draw, rounded corners=2pt, align=center, inner sep=2pt, minimum height=7.5mm},
      mod/.style={box, fill=blue!7, draw=blue!50!black},
      io/.style={box, draw=none, fill=gray!15}]
    \node[io] (d) {Skill\\document};
    \node[mod, right=of d] (ed) {Fine-tuned\\encoder};
    \node[mod, right=of ed] (de) {Doc.-side exp.\\$d \leftarrow \mathrm{n}((1{-}u)d + u\hat q)$};
    \node[mod, right=of de] (bp) {Background proj.\\$d \leftarrow \mathrm{n}(d - P_{B_D} d)$};
    \node[io, below=5mm of d] (q) {Request};
    \node[mod, right=of q] (eq) {Fine-tuned\\encoder};
    \node[mod, right=of eq] (qe) {Query-side exp.\\$q \leftarrow \mathrm{n}((1{-}w)q + w\hat h)$};
    \node[box] (ds) at (bp |- qe) {Dense score\\$\langle q, d\rangle$};
    \node[mod, below=3mm of qe] (lex) {Lexical channel: BM25 on\\request $+$ hypothetical skills};
    \node[box] (fu) at (ds |- lex) {Score fusion\\(Eq.~\ref{eq:fusion})};
    \draw[->] (d) -- (ed); \draw[->] (ed) -- (de); \draw[->] (de) -- (bp); \draw[->] (bp) -- (ds);
    \draw[->] (q) -- (eq); \draw[->] (eq) -- (qe); \draw[->] (qe) -- (ds); \draw[->] (ds) -- (fu);
    \draw[->] (q) |- (lex); \draw[->] (lex) -- (fu);
  \end{tikzpicture}
  \caption{The five modules of SkillRep, where $\hat q$ and $\hat h$ are the mean embeddings of the LLM-written pseudo-requests and hypothetical skills.}
  \label{fig:method}
\end{figure}

Five representation choices make up the space SkillRep covers (Figure~\ref{fig:method}). SkillRep implements each as a module and runs them in one retriever. We stack the unit-norm embeddings of the $N$
documents and the $M$ requests as the rows of $D \in \mathbb{R}^{N \times 1024}$ and
$Q \in \mathbb{R}^{M \times 1024}$, and write $d$ and $q$ for a single document and request
embedding, $P_V = V^{\!\top}V$ for the projector onto the row space of an orthonormal $V$, and
$\mathrm{n}(\cdot)$ for L2 normalization.

\subsection{Encoder fine-tuning}
\label{sec:ft}

We fine-tune Qwen3-Embedding-0.6B~\cite{zhang2025qwen3emb} contrastively on $38{,}308$
request--skill pairs that we synthesize. For each training skill, an LLM writes a request for
which the skill is the right tool, and the request is paired with the skill's body. The training skills are drawn from the non-gold documents of UniPool
(Section~\ref{sec:setup}), and no test request or gold skill has a near duplicate in the
training data (Appendix~\ref{app:ft}).

\subsection{Expansion on both sides}
\label{sec:expansion}

In the two \emph{expansion} modules, an LLM writes text for one side in the style of the
other, and that side's embedding is moved toward it. The query-side module follows
HyDE~\cite{gao2023hyde}. Given the request, an LLM writes two short texts describing a skill
that would solve it (one stating the skill, one describing step by step what it would do),
which we call \emph{hypothetical skills}, and the request embedding moves toward their mean
with weight $w$, which supplies words that the request does not contain. The document-side module
is its mirror, following doc2query~\cite{nogueira2019doc2query}. The LLM writes task
requests for which the skill is the right tool (one stating a goal, one a concrete
problem needing the skill's steps), and the document embedding moves toward their mean
with weight $u$.

\subsection{Background projection}
\label{sec:operators}

Two skill documents can look similar merely because they share a domain or the library's
template wording, and this similarity is the same for the right document and the wrong one.
In the embedding it concentrates in a few directions shared across the library, which we call
the documents' background $B_D$. Background projection subtracts each document's component
along $B_D$ and rescales the result to unit length, $d \leftarrow \mathrm{n}(P_{B_D^\perp}d)$.
The subtraction removes the score $\langle q, P_{B_D}d\rangle$ that the background adds and
leaves the content part $P_{B_D^\perp}d$ unchanged, and the rescaling then enlarges the content part
to unit length, most for documents with a large background.
Requests need no projection because the stored documents lie in $B_D^\perp$, where
$\langle P_{B_D^\perp}q,\, P_{B_D^\perp}d\rangle = \langle q,\, P_{B_D^\perp}d\rangle$, so projecting a request leaves its ranking unchanged.

$B_D$ is the top two right singular vectors of the document matrix after document-side expansion,
which we do not center so that they include the direction all documents share. This is close to the standard approach of
removing the mean and the top principal directions~\cite{mu2018allbutthetop}. The next few
singular vectors also hold background, but mixed with discriminative information, so removing
them would hurt retrieval (Appendix~\ref{app:bgdims}). Background projection therefore adds a
second, narrower step. Following SkillSight~\cite{xiao2026skillsight}, it removes two
directions taken from the library's most generic words, those with the lowest inverse document frequency (IDF), and estimated on the documents after the
first step. This step removes little of each document's length, so it is
subtracted without rescaling, which makes no difference to retrieval (Appendix~\ref{app:bgdims}).
Both steps are computed once over the library, so the projection adds no time at retrieval.

\subsection{Lexical channel}
\label{sec:fusion-main}

The modules above operate on the embedding, so they can use only what the embedding has
kept. The fifth module works on the text itself. Running a sparse retriever alongside a dense
one is standard practice, usually justified as covering exact matches. The lexical channel runs
BM25~\cite{robertson2009bm25}, which scores every document in the library by length-normalized
term overlap, independently of any vector. It searches with the request together with the
hypothetical skills of Section~\ref{sec:expansion}, so query-side expansion reaches both channels.

Skill retrievers that combine a lexical and a dense channel usually fuse them by
reciprocal rank fusion
(RRF)~\cite{cormack2009rrf,li2025skillflow,zheng2026skillselectserve,liu2026wild,ding2026riskexposure}.
However, RRF discards the scores, so a gold document that the dense channel ranks first by a wide margin
looks the same as one it ranks first by a narrow margin, although this margin decides whether
the lexical channel should override the dense ranking (Appendix~\ref{sec:fusion}). We instead
standardize each channel's scores against their distribution over the candidates it
returned, using the z-score normalization of metasearch~\cite{montague2001normalization}, and
add them linearly as in CombSUM~\cite{fox1994combination} and convex-combination hybrid
retrieval~\cite{bruch2023fusion}:
\begin{equation}
f(d) = \frac{s_{\text{dense}}(d) - \mu_{\text{dense}}}{\sigma_{\text{dense}}}
     + \alpha \cdot \frac{s_{\text{bm25}}(d) - \mu_{\text{bm25}}}{\sigma_{\text{bm25}}},
\label{eq:fusion}
\end{equation}
where $\mu$ and $\sigma$ are the mean and standard deviation of that channel's top-100
scores \emph{for this request}, and a document that a channel does not return among its top 100
receives no score from that channel.

\section{Experimental Setup}
\label{sec:setup}

\paragraph{Data.} \textbf{SRA-Bench}~\cite{su2026sra}, our primary benchmark, contributes
5{,}400 requests from six source datasets and an official pool of 26{,}262 skills.
\textbf{SkillsBench}~\cite{li2026skillsbench} is evaluated with the retrieval version
released with SkillRouter~\cite{zheng2026skillrouter}, using its 75 tasks, gold labels and
scoring script, against an \texttt{easy} pool of 78{,}361 skills and a \texttt{hard} pool of 79{,}141 (\texttt{easy} plus 780
adversarial distractors). To evaluate SRA-Bench's requests on a library like the one a deployed agent
would search, we also build \textbf{UniPool}. We start from a 110{,}532-skill snapshot of
SkillCorpus~\cite{wang2026skillcorpus}, skills crawled from 62 public registries, inject
SRA-Bench's gold skills, and remove every pool skill within cosine $0.96$ of a gold skill, as
well as those within $0.88$ that share its name or a 300-character passage. The final pool holds 111{,}247 skills. Unlike SkillsBench's
\texttt{easy} pool ($6.3\%$) and SRA-Bench's official pool ($2.0\%$), UniPool contains no
near-exact duplicates (cosine ${\ge}\,0.999$). We use SkillsBench to compare with published
systems, and SRA-Bench's requests on UniPool for all analyses.

\paragraph{Implementation.} Both expansion modules use Qwen3.6-27B~\cite{yang2025qwen3} at temperature $0.3$, with
two texts per request or per skill. Hypothetical skills are encoded as documents
and pseudo-requests as queries. We set $w{=}0.3$ and $u{=}0.2$. Appendix~\ref{app:wu} shows how
the two weights interact and how well they transfer to other pools. The dense channel indexes
the skill body without \texttt{name} and \texttt{description}, while the lexical channel indexes all three, with BM25 parameters $k_1{=}1.5$ and $b{=}0.75$, fused at $\alpha{=}0.3$.
\textbf{All settings are chosen on UniPool and kept the same on every benchmark.}

\paragraph{Metrics.} We report R@10, MRR@10, and Hit@1, macro-averaged
over SRA-Bench's six subsets. R@10 is the share of a request's gold skills found in its top ten, averaged over requests. MRR@10 averages over requests the reciprocal rank $1/k$ of the first
gold document at rank $k \le 10$, and $0$ if none is in the top ten. We write RR for one
request's reciprocal rank, $\times 100$.

\section{Results}
\label{sec:results}

\subsection{Main results}
\label{sec:mainres}

\begin{table}[t]
    \caption{Results on SRA-Bench (5{,}400 requests, macro averages). $\star$: our runs of each method's released code on the same requests, pools and metrics. Other rows are as published.}
  \label{tab:sra}
  \small
  \setlength{\tabcolsep}{3.5pt}
  \begin{tabular}{@{}llrrr@{}}
    \toprule
    Method & Params & R@10 & MRR@10 & Hit@1\\
    \midrule
    \multicolumn{5}{@{}l}{\emph{Official pool (26{,}262)}}\\
    BM25~\cite{robertson2009bm25}$^\star$ & --- & 69.42 & 54.88 & 44.61\\
    Qwen3-Emb., untuned~\cite{zhang2025qwen3emb}$^\star$ & 0.6B & 68.57 & 57.58 & 48.96\\
    Capability Pages~\cite{wang2026capability} & 8B & 76.72 & 61.72 & ---\\
    Field-aware~\cite{goulart2026fieldaware} & 0.6B & 83.78 & --- & ---\\
        SkillSight~\cite{xiao2026skillsight}$^\star$ & 0.6B & 83.91 & 70.77 & 61.40\\
    SkillRouter-Emb.~\cite{zheng2026skillrouter}$^\star$ & 0.6B & 57.20 & 43.02 & 33.89\\
    Skill2Query~\cite{ding2026skill2query}$^\star$ & 0.6B & 59.19 & 44.97 & 35.26\\
    SkillRet-Emb.~\cite{cho2026skillret}$^{\star}$ & 0.6B & 76.84 & 63.04 & 53.13\\
    SkillReason~\cite{jiang2026skillreason}$^\star$ & 0.6B & 88.18 & 72.19 & 60.85\\
    SkillRep w/o query-side exp. & 0.6B & 89.56 & 74.12 & 64.09\\
    \textbf{SkillRep} & 0.6B & \textbf{92.35} & \textbf{79.21} & \textbf{70.00}\\
    \addlinespace
    \multicolumn{5}{@{}l}{\emph{UniPool (111{,}247)}}\\
    BM25~\cite{robertson2009bm25}$^\star$ & --- & 62.21 & 49.82 & 40.75\\
        SkillSight~\cite{xiao2026skillsight}$^\star$ & 0.6B & 79.55 & 67.67 & 58.66\\
    SkillRouter-Emb.~\cite{zheng2026skillrouter}$^\star$ & 0.6B & 44.86 & 33.76 & 26.29\\
    SkillRet-Emb.~\cite{cho2026skillret}$^{\star}$ & 0.6B & 66.60 & 56.30 & 48.11\\
    SkillReason~\cite{jiang2026skillreason}$^\star$ & 0.6B & 82.87 & 67.30 & 56.60\\
    SkillRep w/o query-side exp. & 0.6B & 85.47 & 70.72 & 60.97\\
    \textbf{SkillRep} & 0.6B & \textbf{89.59} & \textbf{76.46} & \textbf{67.18}\\
    \bottomrule
  \end{tabular}
\end{table}

\paragraph{SRA-Bench.} On the same pools and requests, SkillRep leads the strongest
retriever of its size on every metric and on both pools (Table~\ref{tab:sra}). It leads SkillReason-0.6B
on the official pool by $+7.02$ MRR@10 and $+9.15$ Hit@1, and SkillSight on UniPool by
$+8.79$ and $+8.52$. Without query-side expansion, and so without any LLM call at retrieval
time, it still leads by $1.93$ and $3.05$ MRR@10.
SkillRouter's encoder, which Skill2Query also runs on, transfers poorly to SRA-Bench and scores
below its untuned base with its official instruction.

\begin{table}[t]
  \caption{Results on SkillsBench (75 requests, all 0.6B). Rows without $\star$, other than SkillRep, are copied from~\cite{jiang2026skillreason}, which reports only R@10. $\star$: as in Table~\ref{tab:sra}.}
  \label{tab:sb}
  \small
  \setlength{\tabcolsep}{2.4pt}
  \begin{tabular}{@{}lrrrrrr@{}}
    \toprule
    & \multicolumn{3}{c}{\texttt{easy}} & \multicolumn{3}{c}{\texttt{hard}}\\
    \cmidrule(lr){2-4}\cmidrule(l){5-7}
    Method & R@10 & MRR & Hit@1 & R@10 & MRR & Hit@1\\
    \midrule
    Qwen3-Emb.~\cite{zhang2025qwen3emb} & 61.04 & --- & --- & 57.42 & --- & ---\\
    SkillReason & 67.95 & --- & --- & 62.60 & --- & ---\\
    SkillRouter-Emb.~\cite{zheng2026skillrouter} & 67.96 & --- & --- & 66.07 & --- & ---\\
    Field-aware~\cite{goulart2026fieldaware} & 68.73 & --- & --- & 65.68 & --- & ---\\
    SkillReason$^\star$ & 67.85 & 70.51 & 65.33 & 62.83 & 63.70 & 54.67\\
    SkillSight~\cite{xiao2026skillsight}$^\star$ & 70.13 & 71.45 & 65.33 & 66.70 & 67.36 & 60.00\\
    SkillRet-Emb.~\cite{cho2026skillret}$^{\star}$ & \textbf{72.92} & 70.27 & 64.00 & 64.67 & 64.49 & 56.00\\
    SkillRep w/o query-side exp. & 70.95 & 74.72 & 68.00 & 68.64 & 73.08 & 66.67\\
    \textbf{SkillRep} & 72.07 & \textbf{76.63} & \textbf{72.00} & \textbf{70.10} & \textbf{74.61} & \textbf{69.33}\\
    \bottomrule
  \end{tabular}
\end{table}

\paragraph{SkillsBench.} SkillRep is best on five of the six metrics (Table~\ref{tab:sb}). On
\texttt{easy} R@10, SkillRet's encoder is $0.84$ higher, less than one request's worth on 75
requests, while SkillRep leads it by $6.36$ and $10.12$ MRR@10 on the two tiers. Our runs of
SkillReason-0.6B reproduce its published R@10 within $0.23$ on both tiers, and SkillRep leads
them by $6.12$ and $10.91$ MRR@10.

\paragraph{SkillRet.} SkillRet's requests are synthesized by an LLM from its skills~\cite{cho2026skillret},
so they stay close to the wording of their skill and state more of it than SRA-Bench's requests
do (Appendix~\ref{app:wu}). On its test set (Table~\ref{tab:skillret}), SkillRep without
query-side expansion outperforms every method other than SkillRet's own encoder on every metric.
That encoder, fine-tuned on SkillRet's synthetic training split, leads there but falls $16.17$
and $20.16$ MRR@10 below SkillRep on the two SRA-Bench pools (Table~\ref{tab:sra}).

\begin{table}[t]
  \caption{Results on the SkillRet test set (6{,}006 skills, 4{,}392 requests). All rows are our runs of each method's released code, all 0.6B except BM25. $\dagger$: fine-tuned on SkillRet's training split.}
  \label{tab:skillret}
  \small
  \setlength{\tabcolsep}{4pt}
  \begin{tabular}{@{}lrrr@{}}
    \toprule
    Method & R@10 & MRR@10 & Hit@1\\
    \midrule
    BM25~\cite{robertson2009bm25} & 66.14 & 64.42 & 56.24\\
    Qwen3-Emb., untuned~\cite{zhang2025qwen3emb} & 64.96 & 66.95 & 58.52\\
    SkillSight~\cite{xiao2026skillsight} & 74.10 & 75.50 & 68.33\\
    SkillRouter-Emb.~\cite{zheng2026skillrouter} & 77.24 & 80.84 & 74.34\\
    SkillReason~\cite{jiang2026skillreason} & 77.79 & 81.76 & 75.48\\
    SkillRet-Emb.~\cite{cho2026skillret}$^\dagger$ & \textbf{86.94} & \textbf{84.68} & \textbf{79.12}\\
    SkillRep w/o query-side exp. & 80.40 & 82.53 & 76.39\\
    SkillRep & 80.07 & 81.52 & 74.68\\
    \bottomrule
  \end{tabular}
\end{table}

\subsection{Ablation}
\label{sec:compete}

In Table~\ref{tab:marginal}, we measure each module twice on both encoders, once added \emph{alone} to the encoder with every other module switched off, and
once removed \emph{last} from the full system, which gives what it still adds once the others are
in place. Fine-tuning is the largest either way, $+12.94$ alone and
$+8.79$ last. Among the four training-free modules on the fine-tuned encoder, query-side
expansion and the lexical channel are the largest and background projection the smallest. The
four overlap, as their individual gains sum to $+15.33$ but all four
together give $+13.15$.

The untuned encoder behaves differently. Background projection is worth $+5.55$
alone, in line with the large gains SkillSight reports for training-free
calibration~\cite{xiao2026skillsight}. Query-side expansion falls to $+1.40$, and the four
modules complement rather than overlap ($+16.56$ summed against $+17.30$ together).
The encoder thus determines not only the starting point but also what each
training-free module is worth. Section~\ref{sec:interact} explains these changes by the part of the gap
that each module narrows.

\begin{table}[t]
  \caption{Ablation of each module, added \emph{alone} to the encoder or removed \emph{last} from the full system (UniPool, macro MRR@10). The lexical channel searches with the hypothetical skills only when query-side expansion is on.}
  \label{tab:marginal}
  \small
  \setlength{\tabcolsep}{3.5pt}
  \begin{tabular}{@{}lrrrr@{}}
    \toprule
    Base encoder & \multicolumn{2}{c}{Fine-tuned} & \multicolumn{2}{c}{Untuned}\\
    \cmidrule(lr){2-3}\cmidrule(l){4-5}
    Module & Alone & Last & Alone & Last\\
    \midrule
    Encoder fine-tuning & & & $+$12.94 & $+$8.79\\
    Query-side exp. & $+$6.28 & $+$5.74 & $+$1.40 & $+$4.38\\
    Doc.-side exp. & $+$3.10 & $+$1.07 & $+$4.80 & $+$2.88\\
    Background proj. & $+$1.63 & $+$0.63 & $+$5.55 & $+$4.82\\
    Lexical channel & $+$4.32 & $+$4.25 & $+$4.81 & $+$6.03\\
    \midrule
    Training-free, together & $+$13.15 & & $+$17.30 & \\
    Training-free, sum of \emph{alone} & $+$15.33 & & $+$16.56 & \\
    \bottomrule
  \end{tabular}
\end{table}

\section{Measuring the Gap}
\label{sec:gap}

To explain what each module does and why its value changes with the encoder, we measure the gap between a request and its skill through three
per-request quantities.

\subsection{Background}
\label{sec:decomp}

The first quantity is the background, the component of an embedding that stays the same
whichever request is asked. It does not separate the skill a request needs from the skills competing with it,
yet it occupies part of the embedding's unit length. Such a component can sit on either side
of retrieval, in the embeddings of the skill documents or in those of the requests.

On the documents' side, $B_D$ is the top two right singular vectors of $D$, as in background
projection (Section~\ref{sec:operators}), and the analyses below hold for one to six directions
(Appendix~\ref{app:bgdims}). On the requests' side, $B_Q$ is the direction that requests share,
namely the top right singular vector of the matrix of the requests an LLM writes for the library's
skills (Section~\ref{sec:expansion}), left uncentered so that it keeps the direction all requests
share. It lies $78^\circ$ from the plane of $B_D$, so the documents' estimate misses it.

Both estimates of the background hinder retrieval. Ranking every document by
$\langle q, P_{B_D} d\rangle$ alone, $89\%$ of documents score above the gold one on the
untuned encoder, where a random score would give $50\%$, so the background actively pushes the
gold document down. $B_Q$ interferes because all requests point the same way along it. On the
fine-tuned encoder, all 5{,}400 SRA-Bench requests have a positive component on $B_Q$, $0.27$
on average, so a document's component along $B_Q$ raises its score for every request alike,
and this raise favors the wrong documents. Among the documents in a request's top ten, the gold
ones have a mean component of $0.004$ on $B_Q$ and the others $0.066$, and ranking the library
by $\langle q, P_{B_Q} d\rangle$ alone, $68\%$ of documents score above the gold one even on the
fine-tuned encoder, worse than the $50\%$ of a random score.

We define the \emph{background gap} of a request as how much more the background adds to the
scores of its competitors than to the score of its gold document. With $C(q)$ the non-gold
documents in the top ten of request $q$ and $d^\ast$ its gold document, the background gap along
$B \in \{B_D, B_Q\}$ is
\begin{equation}
  g_B(q) = 100 \cdot \Bigl(\frac{1}{|C(q)|}\sum_{d \in C(q)} \langle q, P_B d\rangle
  - \langle q, P_B d^\ast\rangle\Bigr),
  \label{eq:bggap}
\end{equation}
which we call the $B_D$ gap and the $B_Q$ gap. A positive gap means that the background favors
the competitors. Conditioning on the top ten does not by itself produce a positive gap, because
the competitors in a top ten score below the gold document on average, so selection alone would
push the gap negative.

\subsection{Unsaid vocabulary}
\label{sec:coverage}

The second quantity is the unsaid vocabulary, the words of the needed skill that the request does
not use, since a request states a task in its own terms while the skill describes how to carry it
out. We measure it per request
by the share of the gold document's content words that the request already expresses, which
we call its \emph{coverage}. Low coverage means that much is left unsaid. We count a gold word
as expressed in three ways. \emph{Lexical} coverage requires the same word in the request. A
dense encoder also matches words by meaning, so \emph{semantic} coverage counts a gold word
when some request word lies within cosine $0.8$ of it, with each word embedded on its own by
the untuned encoder. \emph{Soft} coverage drops the threshold and averages each gold word's
highest cosine to any request word.

\subsection{Dilution}

The third quantity is dilution, the part of what a request needs that is lost when its skill
is represented by a single vector. A skill document covers a whole procedure, with its steps,
cases and caveats, while a request typically concerns one of them. The document's vector is a
compromise among all its parts, so the part that the request needs is mixed with parts that
it does not need.

We therefore measure dilution per request by what the request loses when its document is
one vector. With the gold document's body split at its section headings and each section
encoded on its own, let $s^\ast$ be the section most similar to the request $q$. The \emph{dilution loss}
\begin{equation}
  \ell(q) = \cos(q, s^\ast) - \cos(q, d)
  \label{eq:caploss}
\end{equation}
is how much of the request's similarity to the part it needs is lost in the document vector
$d$. It averages $0.072$ on SRA-Bench. It is nearly uncorrelated with the length of the
document ($r = 0.07$) and with coverage ($-0.06$, $-0.04$ and $+0.01$ on the three
counts), so it measures how much of the document is unrelated to this request rather than
how long the document is.

\section{How the Modules Act on the Gap}
\label{sec:why}
\label{sec:matrix}

Sections~\ref{sec:gapbg}--\ref{sec:gapcap} identify the modules that narrow each quantity, as
summarized in Table~\ref{tab:summary}. Section~\ref{sec:interact} then uses these findings to explain why the modules overlap or complement each other. We count a module as narrowing a quantity if its gain varies with the quantity across requests (Table~\ref{tab:matrix}). The quantities are only weakly correlated with each other (Appendix~\ref{app:separate}), so this test does not mistake one for another. Its partial correlations are small because a request's RR takes only eleven values and does not change for most requests when a module is added, and each quantity explains only part of where a module helps. We therefore also change each quantity directly and check that
retrieval moves as predicted, by projecting a background direction out of the documents, deleting
words from the hypothetical skills, and padding the gold documents with content they do not need.
All analyses use SRA-Bench's 5{,}400 requests on UniPool.

\subsection{Background}
\label{sec:gapbg}
\label{sec:finetune}

\begin{table}[t]
  \caption{Which part of the gap each module narrows (Sections~\ref{sec:gapbg}--\ref{sec:gapcap}).}
  \label{tab:summary}
  \small
  \setlength{\tabcolsep}{4pt}
  \begin{tabular}{@{}ll@{}}
    \toprule
    Module & Part of the gap it narrows\\
    \midrule
    Encoder fine-tuning & background ($B_D$, $B_Q$)\\
    Query-side exp. & unsaid vocabulary, $B_Q$ after fine-tuning\\
    Doc.-side exp. & background ($B_Q$)\\
    Background proj. & background ($B_D$, $B_Q$)\\
    Lexical channel & dilution\\
    \bottomrule
  \end{tabular}
\end{table}

\begin{table}[t]
  \caption{Partial correlation $\rho$ (Eq.~\ref{eq:partialrho}) of each module's gain, when added alone, with each measure (UniPool). Bold: the module gains more where that part of the gap is larger, with $|\rho| \ge 0.10$.}
  \label{tab:matrix}
  \small
  \setlength{\tabcolsep}{2.2pt}
  \begin{tabular}{@{}lrrrrrr@{}}
    \toprule
    & \multicolumn{3}{c}{Coverage} & \multicolumn{2}{c}{Gap} & \\
    \cmidrule(lr){2-4}\cmidrule(lr){5-6}
    Module & Lex. & Sem. & Soft & $B_D$ & $B_Q$ & Dilution\\
    \midrule
    Fine-tuning & $+$.03 & $+$.03 & $+$.05 & $-$.09 & $-$.08 & $-$.18\\
    Query-side exp. & $\mathbf{-.12}$ & $\mathbf{-.17}$ & $\mathbf{-.15}$ & $-$.03 & $\mathbf{+.20}$ & $-$.08\\
    Doc.-side exp. & $+$.08 & $+$.07 & $+$.06 & $-$.08 & $\mathbf{+.13}$ & $+$.02\\
    Background proj. & $+$.04 & $+$.06 & $+$.04 & $\mathbf{+.10}$ & $\mathbf{+.11}$ & $-$.02\\
    Lexical channel & $+$.04 & $-$.01 & $+$.00 & $-$.04 & $+$.03 & $\mathbf{+.10}$\\
    \bottomrule
  \end{tabular}
\end{table}

\begin{table}[t]
    \caption{The background gap of each encoder and the change each module makes to it (UniPool, $\times 100$), with $B$ estimated in each encoder's own space. Bold: changes of at least $0.25$.}
  \label{tab:matrixbg}
  \small
  \setlength{\tabcolsep}{4pt}
    \begin{tabular}{@{}lrr@{}}
    \toprule
    & $B_D$ & $B_Q$\\
    \midrule
    Untuned encoder & 6.87 & 6.79\\
    \;+ fine-tuning & $\mathbf{-6.59}$ & $\mathbf{-5.02}$\\
    \midrule
    Fine-tuned encoder & 0.28 & 1.77\\
    \;+ query-side exp. & $-$0.14 & $\mathbf{-0.99}$\\
    \;+ doc.-side exp. & $-$0.02 & $\mathbf{-0.56}$\\
    \;+ background proj. & $\mathbf{-0.28}$ & $\mathbf{-0.68}$\\
    \;+ lexical channel & $-$0.03 & $-$0.12\\
    \bottomrule
  \end{tabular}
\end{table}

As for the other quantities, whether a module's gain varies with the background can be tested for every module (Table~\ref{tab:matrix}). The background, however, is measured on the embeddings, which most modules change, so for it we can also test more directly whether adding a module narrows it: Table~\ref{tab:matrixbg} reports the background gap (Eq.~\ref{eq:bggap}) and how each module changes it.

\textit{(1) Fine-tuning} narrows the background gap most. According to Table~\ref{tab:matrixbg}, it
narrows it by $6.59$ along $B_D$ and $5.02$ along $B_Q$, because it shortens the background. With $B_D$ estimated in each encoder's own space, a request's norm on $B_D$ falls from $0.348$ to $0.115$.

\textit{(2) Query-side expansion} narrows the $B_Q$ gap most, by $0.99$, because a request
mixed with a hypothetical skill has a smaller component along $B_Q$.

\textit{(3) Document-side expansion} leaves the $B_D$ gap unchanged and narrows the $B_Q$
gap by $0.56$. Its effect on $B_Q$ does not come from supplying unsaid vocabulary from the documents' side but has a simpler cause. Competitors in the top
ten have a larger component along $B_Q$ than the gold document (the $B_Q$ gap of $1.77$ in
Table~\ref{tab:matrixbg}), and mixing pseudo-requests, which all have a large component along
$B_Q$, into every document evens out this difference. To test this explanation, we run two
controls on the fine-tuned encoder. Projecting $B_Q$ out of the documents, with no
pseudo-requests mixed in, adds $2.96$ MRR@10, nearly as much as the module's $3.10$. Mixing
the same library-wide average of pseudo-requests into every document, with nothing specific to
any skill, adds $1.67$. Of document-side expansion's $3.10$, $1.67$ thus comes from a component
that every document receives alike, and the rest requires each skill's own pseudo-requests.

\textit{(4) Background projection} removes the remaining component along $B_D$ and narrows the $B_Q$ gap as well.

Consistent with these results, Table~\ref{tab:matrix} shows that query-side expansion, document-side expansion and background projection gain more where the $B_Q$ gap is larger, and background projection also where the $B_D$ gap is larger.

\subsection{Unsaid vocabulary}
\label{sec:gapvocab}
\label{sec:bridgemech}

We measure the unsaid vocabulary by the three counts of coverage
(Section~\ref{sec:coverage}) and test whether a module's gain correlates with them.
A plain correlation would mislead in two ways. The baseline RR rises with coverage on every
count, so a gain can shrink toward high coverage merely because there is less left to gain, and
coverage is slightly correlated with the other measures (Appendix~\ref{app:separate}), so part
of its correlation with a gain could belong to them. The partial correlation $\rho$ removes both
effects. For
request $i$, let $c_i$ be its value of one measure, $g_i$ the change that the module makes to
its RR, and $z_i$ its RR on the encoder that the module is added to, together with the other
measures. With $\hat c(\cdot)$ and $\hat g(\cdot)$ the least-squares linear fits of $c$ and $g$
on $z$,
\begin{equation}
  \rho = \operatorname{corr}_i\bigl(c_i - \hat c(z_i),\; g_i - \hat g(z_i)\bigr),
  \label{eq:partialrho}
\end{equation}
over all $5{,}400$ requests. A negative $\rho$ with coverage means the module helps more where
the request leaves more unsaid, beyond what the lower starting point alone would give. As shown in Table~\ref{tab:matrix}, only query-side expansion's gain falls with coverage, on every count ($\rho = -0.12$,
$-0.17$ and $-0.15$). The other modules' $\rho$ lie between $-0.01$ and $+0.08$. This is what we expect if query-side
expansion supplies the vocabulary that a request leaves unsaid.

To find the mechanism by which query-side expansion supplies the unsaid vocabulary, we delete from the hypothetical skills the gold-document words that the request lacks, which lowers the module's gain by $3.12$ MRR@10. Deleting the same number of other words,
each matched to a deleted word by IDF, lowers it by only $0.08$.
Keeping only those gold words and nothing else of the hypothetical skills raises the gain by a
further $4.27$. Both edits use
the gold document, so they are probes rather than a method, but they show that \textbf{the gain is
carried by the gold-document words that the LLM supplies.}

\subsection{Dilution}
\label{sec:gapcap}
\label{sec:immune}

\begin{figure}[t]
  \centering
  \begin{tikzpicture}
  \begin{axis}[width=\columnwidth, height=3.9cm, font=\footnotesize,
      xmin=0.7, xmax=5.3, ymin=40, ymax=80, xtick={1,...,5},
      xticklabels={{$F_1$\\[-2pt].000},{$F_2$\\[-2pt].039},{$F_3$\\[-2pt].065},{$F_4$\\[-2pt].096},{$F_5$\\[-2pt].162}},
      xticklabel style={align=center}, ylabel={RR}, ylabel style={yshift=-4pt},
      legend columns=3, legend style={at={(0.5,1.03)}, anchor=south, draw=none, font=\scriptsize,
        /tikz/every even column/.append style={column sep=4pt}},
      every axis plot/.append style={thick, mark size=1.6pt}]
    \addplot[color=red!75!black, mark=*] coordinates {(1,77.3) (2,68.8) (3,64.1) (4,59.2) (5,56.1)};
    \addplot[color=blue!70!black, mark=square*] coordinates {(1,70.9) (2,61.0) (3,58.8) (4,57.6) (5,60.3)};
    \addplot[color=black!60, mark=triangle*] coordinates {(1,57.7) (2,47.0) (3,44.1) (4,52.3) (5,63.0)};
    \legend{One vector ($\rho{=}{-}.168$), Sections ($-.050$), BM25 ($+.074$)}
  \end{axis}
  \end{tikzpicture}
  \caption{RR of each retriever by fifths of the gold document's dilution loss $\ell$, with the mean $\ell$ below each fifth (UniPool, micro averages). $\rho$: partial correlation of RR with $\ell$ (Section~\ref{sec:gapcap}).}
  \label{fig:dilution}
\end{figure}
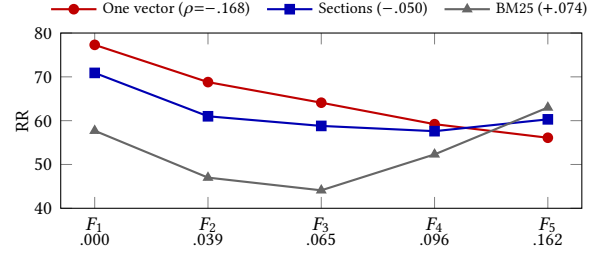

We measure dilution by the dilution loss of Eq.~\ref{eq:caploss}. If dilution is a weakness
of representing a document by one vector, retrieval that keeps one vector per document
should lose accuracy as the gold document's dilution loss grows, and retrieval that sees
the parts of the document should not. We test this with three separate retrievers: (i) the fine-tuned encoder, which keeps one vector per body; (ii) the same encoder indexing each section
separately, prefixed with the skill's \texttt{name} and \texttt{description}, and scoring a document by its
best section, as passage-level retrieval does; and (iii) BM25 with the request alone.
Figure~\ref{fig:dilution} splits the requests into fifths by the gold document's dilution
loss. Only the one-vector retriever is hurt. Its RR falls from $77.3$ on the least diluted fifth
to $56.1$ on the most diluted, and the partial correlation is $-0.168$ once
coverage, length and the number of sections are held fixed. Neither the section-level
retriever ($-0.050$) nor BM25 ($+0.074$) loses accuracy with dilution, and BM25 is at its best
on the most diluted fifth. \textbf{What the dilution loss measures is therefore specific to keeping
one vector per document, not a property of the document that makes it hard to retrieve by
any means.}

Because $\ell$ subtracts the one-vector score $\cos(q,d)$ itself, part of this correlation is
built into the measure. To show that dilution itself costs the single vector accuracy, we add
dilution directly, appending to each gold document sections that it does not need and leaving
every request unchanged. We pad with three kinds of sections. Sections from other skills add
new content on another topic, the library's most generic sections add new content on no
particular topic, and copies of the gold document's own least relevant section add length but
no new content. If dilution rather than length is the cause,
the one-vector retriever should lose more where $\ell$ rises more, and section-level
retrieval should not lose. According to Table~\ref{tab:padding}, sections from other skills raise $\ell$ most and cost the
one-vector retriever up to $9.68$ MRR@10, generic sections raise it less and
cost up to $5.17$, and own copies slightly lower $\ell$ (from $.073$ to $.065$) and cost at most $1.38$. Section-level retrieval does not
change. BM25 falls by similar amounts under all three, as a longer document would.
\textbf{Content that a request does not need thus hurts the single vector by diluting it, not
by lengthening the document.}

\begin{table}[t]
  \caption{Effect of padding the gold documents with content the requests do not need (UniPool, macro MRR@10, change from no padding). \emph{Other}: $L$ sections drawn at random from other skills. \emph{Generic}: $L$ of the library's most template-like sections. \emph{Own}: $L$ copies of the gold document's least relevant section. $\ell$: mean dilution loss.}
  \label{tab:padding}
  \small
  \setlength{\tabcolsep}{3.5pt}
  \begin{tabular}{@{}lrrrrr@{}}
    \toprule
    Padding & $L$ & $\ell$ & One vector & Sections & BM25\\
    \midrule
    None & 0 & .073 & 63.43 & 60.67 & 49.82\\
    \midrule
    Other & 2 & .091 & $-$7.03 & $+$0.08 & $-$1.32\\
          & 8 & .108 & $-$9.68 & $+$0.28 & $-$6.19\\
    Generic & 2 & .074 & $-$2.72 & $+$0.17 & $-$1.59\\
          & 8 & .083 & $-$5.17 & $+$0.37 & $-$6.26\\
    Own   & 2 & .068 & $-$0.70 & $+$0.04 & $-$1.50\\
          & 8 & .065 & $-$1.38 & $\phantom{+}$0.00 & $-$5.72\\
    \bottomrule
  \end{tabular}
\end{table}

We then identify the module that recovers what dilution costs, by testing whether its gain rises with the dilution loss across requests. Only the lexical channel's gain does
($\rho = +0.10$ in Table~\ref{tab:matrix}, searching with the request alone).

To find the mechanism by which the lexical channel recovers what dilution costs, we compare
the requests whose gold document it moves into the top ten with those whose gold document it
pushes out. What separates the two groups is how many more content words a request shares
with the gold document than with its strongest dense competitor, $3.04$ against $0.21$, about
evenly split between rare and common words. These are the words that dilution hides from the
single vector. Deleting them from the section of the gold document that best matches the
request lowers that section's similarity to the request by $0.050$, but lowers the whole
document's by only $0.007$. Deleting as many other words of similar IDF lowers neither. The
channel thus \textbf{helps by counting the words a request shares with the document, which the
document's single vector dilutes.} Appendix~\ref{sec:fusion} shows how this mechanism
decides how the channel should be fused.

\subsection{Why the modules overlap}
\label{sec:interact}

\begin{table}[t]
  \caption{How much document-side expansion and background projection narrow the $B_Q$ gap of
  Table~\ref{tab:matrixbg} and gain in MRR@10, added alone or after query-side expansion
  (UniPool, dense channel only). The first row is query-side expansion's own
  narrowing.}
  \label{tab:overlap}
  \small
  \setlength{\tabcolsep}{4pt}
  \begin{tabular}{@{}lrrrr@{}}
    \toprule
    & \multicolumn{2}{c}{Untuned} & \multicolumn{2}{c}{Fine-tuned}\\
    \cmidrule(lr){2-3}\cmidrule(l){4-5}
    & Alone & After QE & Alone & After QE\\
    \midrule
    \multicolumn{5}{@{}l}{\emph{$B_Q$ gap narrowed by}}\\
    \;Query-side exp. & $-$0.19 & & 0.99 & \\
    \;Doc.-side exp. & 1.65 & 1.58 & 0.56 & 0.25\\
    \;Background proj. & 6.57 & 6.82 & 0.68 & 0.29\\
    \addlinespace
    \multicolumn{5}{@{}l}{\emph{MRR@10 gain}}\\
    \;Doc.-side exp. & $+$4.80 & $+$5.02 & $+$3.10 & $+$2.10\\
    \;Background proj. & $+$5.55 & $+$6.47 & $+$1.63 & $+$0.96\\
    \bottomrule
  \end{tabular}
\end{table}

Table~\ref{tab:marginal} shows that the four training-free modules overlap on the fine-tuned
encoder but complement each other on the untuned one, and that fine-tuning weakens background
projection and strengthens query-side expansion. Given which part of the gap each module
narrows (Table~\ref{tab:summary}), one principle explains all of this:
\textbf{modules that narrow the same part of the gap overlap, and modules that narrow different
parts complement each other.}

\paragraph{Modules that share a part of the gap overlap.} Document-side expansion and
background projection narrow only the background, which fine-tuning and query-side expansion
have already narrowed. After query-side expansion, both narrow the $B_Q$ gap less and gain
correspondingly less (Table~\ref{tab:overlap}). The three modules overlap by $2.11$ MRR@10,
nearly all of the $2.18$ by which the four overlap in Table~\ref{tab:marginal}. Fine-tuning and
background projection overlap in the same way along $B_D$. Fine-tuning has already shortened
$B_D$ (Section~\ref{sec:gapbg}), so projecting out the remainder adds $+1.63$ MRR@10 instead of
$+5.55$, for any number of directions (Appendix~\ref{app:bgdims}).

\paragraph{Modules with a part of the gap of their own keep their gain.} The lexical channel,
the only module that acts on dilution, loses almost nothing when added last ($+4.32$ alone and $+4.25$ last in Table~\ref{tab:marginal}), and query-side expansion, the only one that supplies
unsaid vocabulary, keeps most of its gain. On the untuned encoder, query-side expansion does not narrow the $B_Q$ gap
(Table~\ref{tab:overlap}), so it does not overlap with the other modules. With both
backgrounds projected out of the documents, its gain rises from $+1.40$ to $+2.41$, background
projection gains more after it than alone ($+6.47$ against $+5.55$), and document-side expansion
gains about as much ($+5.02$ against $+4.80$).

\paragraph{Fine-tuning moves query-side expansion onto $B_Q$.} Query-side expansion narrows the $B_Q$ gap only after fine-tuning, because fine-tuning changes
which embeddings lean along $B_Q$. On the untuned encoder, $B_Q$
nearly coincides with the direction that all documents share (cosine $0.95$ with their mean),
and documents and hypothetical skills lean along it as much as requests do, so mixing a
hypothetical skill into a request does not reduce the request's component on it. Fine-tuning spreads
the documents out and turns $B_Q$ into a direction along which only requests lean (cosine
$0.22$), so the same mixing now reduces it. This gives query-side expansion a second source of
gain, as projecting $B_Q$ out first lowers its gain from $+6.28$ to $+4.27$, and, as the principle predicts, makes it
overlap with document-side expansion and background projection.

\section{Conclusion}
\label{sec:conclusion}

Requests and skills are written for different purposes and in different words, and prior
retrievers each narrow this gap with one or two representation choices. We combined encoder
fine-tuning, query-side and document-side expansion, background projection and a lexical channel
as modules of one retriever, SkillRep, which sets a new state of the art on SRA-Bench and
SkillsBench with a 0.6B encoder. The value of each module depends on the others, and
fine-tuning in particular reverses the ranking of the four training-free modules. We explained
this by measuring the gap through three per-request quantities, namely the background, the
unsaid vocabulary and dilution. Each module narrows a particular part of the gap, and modules
that narrow the same part overlap while those that narrow different parts complement each other.
The quantities also show how the modules work. Query-side expansion helps through the gold
document's words that the LLM supplies, and the lexical channel recovers what dilution costs the
single vector by counting the words a request shares with its skill. A module's gain is
therefore best reported together with the encoder beneath it and the part of the gap it narrows.

\bibliographystyle{ACM-Reference-Format}
\bibliography{refs}

\appendix

\section{Encoder fine-tuning}
\label{app:ft}

Each training pair comes with up to ten negatives, seven mined from the untuned encoder's top 50
and three random. Training uses an in-batch InfoNCE
loss~\cite{oord2018cpc} at temperature $0.05$, with negatives shared across eight GPUs, for one
epoch at learning rate $2{\times}10^{-5}$ and maximum length $1{,}024$. The training skills come
from a registry snapshot that overlaps UniPool, so the encoder sees part of the library it later
searches, but none of SRA-Bench's gold skills. Training on the searched library does not appear to explain the
lead. The training skills make up $30\%$ of UniPool but only $13\%$ of SRA-Bench's official
pool, yet SkillRep's lead over SkillSight is about as large on the official pool, $8.44$
against $8.79$ MRR@10 (Table~\ref{tab:sra}). No test request lies within cosine $0.90$ of a
training request, and no gold skill within cosine $0.90$ of a training skill. The closest
pairs reach $0.83$ and $0.77$.

\section{Score fusion against rank fusion}
\label{sec:fusion}

The mechanism of Section~\ref{sec:immune} has a cost as well as a benefit, and the cost decides how the
lexical channel should be fused. Matching shared words rescues the requests whose gold document the dense channel
ranks below a competitor, but it can also displace a gold document that the dense channel
already ranks first. Table~\ref{tab:bm25split} sorts requests by the gold document's dense
margin over its strongest competitor and shows both. Under rank fusion, the channel is worth
$+12.74$ and $+7.88$ RR points in the two fifths where the gold document trails most, and
costs $6$ to $8$ points in the two where it leads. At retrieval time, however, the gold
document is unknown, so the channel cannot be applied to the first group alone. Only the distance between the dense scores is observable. Rank fusion discards this distance, so a better lexical
rank can overturn a dense lead of any size, and averaged over all requests the gains and losses
nearly cancel, and the channel adds only $+1.34$ when added last. Score fusion keeps the distance, so the lexical evidence needed to overturn the
dense ranking grows with the dense margin, so a wide lead is hard to displace, while a narrow one
can be reversed. Under score fusion the channel adds $+4.25$ (Table~\ref{tab:marginal}), which
is why SkillRep fuses the two channels by score (Section~\ref{sec:fusion-main}).

\begin{table}[h!]
    \caption{The lexical channel's value under \emph{rank} fusion, by the gold document's dense margin over its strongest competitor (UniPool, fifths of requests). Gold's percentile: share of the library scored above the gold document.}
  \label{tab:bm25split}
  \small
  \setlength{\tabcolsep}{4pt}
  \begin{tabular}{@{}lrrrrr@{}}
    \toprule
        Gold's margin & $-$.136 & $-$.051 & $-$.008 & $+$.046 & $+$.145\\
    \midrule
    Gold's percentile & 0.244 & 0.006 & 0.001 & 0.000 & 0.000\\
    $\Delta$RR from the channel & \up{12.74} & \up{7.88} & \dn{0.84} & \dn{7.95} & \dn{6.38}\\
    \bottomrule
  \end{tabular}
\end{table}

\begin{table}[h!]
  \caption{Score fusion against rank fusion at $\alpha{=}0.3$ in the full system, other modules unchanged.}
  \label{tab:fusion}
  \small
  \setlength{\tabcolsep}{3.5pt}
  \begin{tabular}{@{}lrrrrrr@{}}
    \toprule
    & \multicolumn{2}{c}{R@10} & \multicolumn{2}{c}{MRR@10} & \multicolumn{2}{c}{Hit@1}\\
    \cmidrule(lr){2-3}\cmidrule(lr){4-5}\cmidrule(l){6-7}
    Pool & rank & score & rank & score & rank & score\\
    \midrule
    SRA official & 92.00 & 92.35 & 76.16 & 79.21 & 65.30 & 70.00\\
    UniPool & 89.52 & 89.59 & 73.56 & 76.46 & 62.39 & 67.18\\
    SkillsB.\ \texttt{easy} & 69.89 & 72.07 & 73.67 & 76.63 & 68.00 & 72.00\\
    SkillsB.\ \texttt{hard} & 68.69 & 70.10 & 73.55 & 74.61 & 68.00 & 69.33\\
    \bottomrule
  \end{tabular}
\end{table}

If score fusion helps by keeping the dense margin, it should mainly change \emph{which document is placed first} rather than which documents are
retrieved, and Table~\ref{tab:fusion} shows that it does. Score fusion is better on every
pool and metric, worth $+3.05$ and $+2.90$ MRR@10
on the two SRA-Bench pools, but there R@10 barely moves ($+0.35$, $+0.07$) while Hit@1
rises by $4.70$ and $4.79$.

\section{The expansion weights and their interactions}
\label{app:wu}
Throughout this appendix $w$ and $u$ weight the dense channel only. The lexical channel
searches with the hypothetical skills at every $w$, so $w{=}0$ turns query-side
expansion off in the dense channel but not in the lexical one, whereas ``SkillRep w/o
query-side exp.'' in Table~\ref{tab:sra} removes it from both.
The two expansion weights overlap in the full system (Table~\ref{tab:wu}). Each row sweeps the
document-side weight $u$ at a fixed query-side weight $w$. As $w$ grows, the best $u$
moves left and the gain it yields shrinks from $1.67$ to $0.76$. Tuning each weight by itself
and combining the winners, $(0.3, 0.3)$, lands $0.52$ MRR@10 below the jointly chosen
$(0.3, 0.2)$.

\begin{table}[h!]
  \caption{MRR@10 of the full system over the grid of expansion weights $(w,u)$ (UniPool, macro averages).}
  \label{tab:wu}
  \small
  \setlength{\tabcolsep}{5pt}
  \begin{tabular}{@{}lrrrrrcr@{}}
    \toprule
    & \multicolumn{5}{c}{document-side weight $u$} & & \\
    \cmidrule(lr){2-6}
    $w$ & 0 & 0.1 & 0.2 & 0.3 & 0.5 & $u^\star$ & gain\\
    \midrule
    0 & 72.09 & 73.11 & 73.55 & 73.76 & 72.81 & 0.3 & $+$1.67\\
    0.3 & 75.39 & 76.17 & 76.46 & 75.94 & 74.46 & 0.2 & $+$1.07\\
    0.5 & 74.27 & 74.76 & 75.03 & 74.76 & 73.20 & 0.2 & $+$0.76\\
    \bottomrule
  \end{tabular}
\end{table}

The weights were chosen on UniPool, so Table~\ref{tab:wutransfer} repeats the grid on the
four other pools. On the official SRA-Bench pool,
$(0.3, 0.2)$ is the best of the fifteen settings. On SkillsBench it is $0.20$ (easy) and
$0.09$ (hard) MRR@10 below the best, differences within the noise of a $75$-request
benchmark, and the best settings there do not agree between the two pools. On SkillRet
(Section~\ref{sec:mainres}) it is $1.56$ below the best, $(0, 0.5)$, and the shape of the
grid follows from that benchmark's requests, which state more of their skill (median
coverage $0.100$, $0.353$ and $0.795$ on the three counts, against $0.046$, $0.236$ and $0.769$ on
SRA-Bench), so query-side expansion has little left to supply in the dense channel and the
best $w$ there is $0$, while document-side expansion helps more the higher $u$ is.

The settings were also chosen on the same SRA-Bench requests they are tested on. To bound what
that choice is worth, we split the requests in half at random, stratified by source dataset,
search the full grid of the weights, the projection (none, one step, two levels), the fusion
(none, rank or score, at six weights) and the indexed granularity on one half, $3{,}150$
settings, and test the winner on the other half. It scores $75.56$ MRR@10 there, $0.60$
below the best setting for that half, against a lead of $7$ to $9$ points over the strongest baselines.

\begin{table}[h!]
  \caption{MRR@10 over the grid of expansion weights $(w,u)$ on the pools not used to choose them. Bold: best per pool. Underlined: the chosen $(0.3, 0.2)$.}
  \label{tab:wutransfer}
  \small
  \setlength{\tabcolsep}{5pt}
  \begin{tabular}{@{}llrrrrr@{}}
    \toprule
    & & \multicolumn{5}{c}{document-side weight $u$}\\
    \cmidrule(lr){3-7}
    Pool & $w$ & 0 & 0.1 & 0.2 & 0.3 & 0.5\\
    \midrule
    SRA official & 0 & 77.01 & 77.32 & 77.67 & 77.84 & 76.75\\
     & 0.3 & 78.95 & 79.15 & \underline{\textbf{79.21}} & 79.13 & 77.99\\
     & 0.5 & 77.59 & 77.88 & 77.81 & 77.80 & 76.84\\
    \midrule
    SkillsBench easy & 0 & 76.04 & 75.89 & 74.86 & 73.84 & 71.25\\
     & 0.3 & 76.02 & 75.86 & \underline{76.63} & 74.97 & 73.30\\
     & 0.5 & 76.23 & \textbf{76.83} & 76.64 & 75.20 & 72.43\\
    \midrule
    SkillsBench hard & 0 & 73.46 & 74.46 & 73.34 & 72.51 & 68.98\\
     & 0.3 & \textbf{74.70} & 73.79 & \underline{74.61} & 72.84 & 70.30\\
     & 0.5 & 74.12 & 74.06 & 74.40 & 72.91 & 68.80\\
    \midrule
    SkillRet & 0 & 79.41 & 80.76 & 81.82 & 82.69 & \textbf{83.08}\\
     & 0.3 & 79.65 & 80.81 & \underline{81.52} & 82.19 & 82.39\\
     & 0.5 & 78.77 & 79.78 & 80.56 & 80.93 & 81.10\\
    \bottomrule
  \end{tabular}
\end{table}

\section{Background projection: controls and dimensionality}
\label{app:bgdims}

\begin{table}[h!]
  \caption{Effect of projecting different directions out of the documents (UniPool, MRR@10 change). \emph{SVD, top $k$}: the top $k$ singular vectors of the document matrix, projected out in one step. \emph{Fine-tuned}: the fine-tuned encoder alone ($63.31$). \emph{Full}: the full system with its projection replaced ($75.82$ without one). \emph{Untuned}: the untuned encoder alone ($50.37$). Random: mean of ten seeds.}
  \label{tab:bgdims}
  \small
  \setlength{\tabcolsep}{4pt}
  \begin{tabular}{@{}lrrr@{}}
    \toprule
    Directions projected out & Fine-tuned & Full & Untuned\\
    \midrule
    2 random & $-$0.00 & --- & ---\\
    4 random & $-$0.01 & --- & ---\\
    \addlinespace
    SVD, top 1 & $+$1.01 & $+$0.11 & $+$4.12\\
    SVD, top 2 ($B_D$) & $+$0.88 & $+$0.23 & $+$5.04\\
    SVD, top 3 & $+$1.63 & $+$0.39 & $+$5.65\\
    SVD, top 4 & $+$2.46 & $+$0.23 & $+$5.54\\
    SVD, top 6 & $+$2.61 & $+$0.08 & $+$5.47\\
    SVD, top 8 & $+$2.30 & $+$0.01 & $+$5.91\\
    Generic words only, 2 & $+$0.81 & $-$0.19 & ---\\
    \addlinespace
    Two-level, $2{+}2$ (ours) & $+$1.63 & $+$0.63 & $+$5.55\\
    \bottomrule
  \end{tabular}
\end{table}

In Table~\ref{tab:bgdims}, random directions of the same number change nothing, so the gain is not a matter of
dimension. On the fine-tuned encoder alone, removing more of the top singular directions improves retrieval up to six. In the full system, where the other modules already narrow part of the background, a single step improves it only up to three, and the two-level projection, which removes two further directions only where they align with generic words, is best. On the untuned encoder every single step of two to eight directions
is worth $+5.0$ to $+5.9$, against at most $+2.6$ on the fine-tuned one.

The background analyses of Section~\ref{sec:gapbg} do not depend on the number of
directions in $B_D$. Taking the top $k$ singular directions for $k$ from $1$ to $6$, the background gap of
Table~\ref{tab:matrixbg} lies between $6.07$ and $7.34$ on the untuned encoder and between
$0.16$ and $1.60$ on the fine-tuned one.

After the first step, a document's squared length along the second step's two directions is only $0.047$ on average, so rescaling after the second step changes MRR@10 by only $-0.02$ in the full system and $+0.10$ with the projection alone.

\section{Separability of the three quantities}
\label{app:separate}

\begin{table}[H]
  \caption{Pearson correlations between the per-request measures (UniPool).}
  \label{tab:separate}
  \small
  \setlength{\tabcolsep}{3pt}
  \begin{tabular}{@{}lrrrrrr@{}}
    \toprule
    & \multicolumn{3}{c}{Coverage} & Gap & Gap & \\
    \cmidrule(lr){2-4}
    & Lex. & Sem. & Soft & $B_D$ & $B_Q$ & Dilution\\
    \midrule
    Cov., lexical & 1 & 0.88 & 0.83 & $-$0.12 & $-$0.12 & $-$0.06\\
    Cov., semantic & & 1 & 0.93 & $-$0.24 & $-$0.22 & $-$0.04\\
    Cov., soft & & & 1 & $-$0.29 & $-$0.23 & 0.01\\
    Gap $B_D$ & & & & 1 & 0.34 & $-$0.13\\
    Gap $B_Q$ & & & & & 1 & 0.04\\
    \bottomrule
  \end{tabular}
\end{table}

The three quantities are only weakly correlated ($|r| \le 0.29$), except that the three
counts of coverage agree with each other ($0.83$ to $0.93$) and the two estimates of the
background share part of their variation ($0.34$). Table~\ref{tab:matrix} therefore controls
for the other measures when it relates a module's gain to one of them.

\section{The three quantities beyond skills}
\label{app:generalir}

\begin{table}[H]
  \caption{The three quantities on skills (SRA-Bench requests on UniPool) and on three BEIR
  collections~\cite{thakur2021beir}, with the untuned Qwen3-Embedding-0.6B, the dense channel only and micro averages.
  \emph{Share of $B_D$}: mean squared length of a document along $B_D$. \emph{Above gold}: share
  of the collection that scores above the gold document on $\langle q, P_{B_D} d\rangle$ alone.
  \emph{Gap}: the background gap along $B_D$ of Table~\ref{tab:matrixbg}. \emph{Projection}:
  MRR@10 change from projecting $B_D$ out. \emph{Request words}: median content words per
  request. \emph{Coverage}: lexical coverage of Section~\ref{sec:coverage} (gold side), and the
  share of a request's content words found in its gold document (request side). $\ell$:
  dilution loss with gold documents cut into 80-word windows.}
  \label{tab:generalir}
  \small
  \setlength{\tabcolsep}{3pt}
  \begin{tabular}{@{}lrrrr@{}}
    \toprule
    & Skills & SciFact & NFCorpus & FiQA\\
    \midrule
    Documents & 111{,}247 & 5{,}183 & 3{,}633 & 57{,}638\\
    Requests & 5{,}400 & 300 & 323 & 648\\
    MRR@10 & 50.15 & 66.64 & 56.36 & 55.30\\
    \midrule
    Share of $B_D$ & .34 & .33 & .34 & .27\\
    Above gold & 89\% & 28\% & 41\% & 38\%\\
    Gap & 6.87 & 0.19 & 0.98 & 0.64\\
    Projection & $+$5.65 & $-$0.76 & $-$0.98 & $-$0.17\\
    \midrule
    Request words & 21 & 8 & 2 & 6\\
    Coverage, gold side & .076 & .049 & .007 & .048\\
    Coverage, request side & .39 & .51 & .23 & .43\\
    \midrule
    Gold words & 504 & 230 & 232 & 191\\
    $\ell$ & .077 & .019 & .026 & $-$.001\\
    \bottomrule
  \end{tabular}
\end{table}

To see how skill retrieval differs from general retrieval, we measure the three quantities on
three BEIR collections~\cite{thakur2021beir}, SciFact, NFCorpus and FiQA, with the untuned
encoder and the dense channel only (Table~\ref{tab:generalir}). The definitions are those of Section~\ref{sec:gap}, except that the dilution loss cuts every gold document into 80-word windows, as BEIR documents have no sections. Two of the quantities set skill retrieval apart. Only in
the skill library does the background score the competitors above the gold document, and
dilution is strongest there. Coverage is not lower for skills, but a skill request shares fewer
of its words with its gold document than most BEIR queries do.

The background takes a similar share of the document embeddings in every collection ($0.27$ to
$0.34$). Ranked by $\langle q, P_{B_D} d\rangle$ alone, however, $89\%$ of the library scores above the
gold skill, against $28$ to $41\%$ of a BEIR collection, fewer than the $50\%$ of a random score. Projecting $B_D$
out accordingly adds $5.65$ MRR@10 on skills and lowers it by $0.17$ to $0.98$ on BEIR.

The dilution loss is $0.077$ on skills, against
$-0.001$ to $0.026$ on BEIR, as a skill document describes a whole procedure and runs to $504$
words on average, against $191$ to $232$.

Coverage on the gold side is higher for skills ($0.076$, against $0.007$ to $0.049$), but only
because a skill request is long, $21$ content words against $2$ to $8$. From the request
side, only $39\%$ of a skill request's content words occur in its gold skill, against $51\%$ and $43\%$ on SciFact and FiQA, since a
task states details that a general skill does not. Only NFCorpus, whose queries are two words
long, is lower.

\end{document}